\documentclass[final,epsfig,epj]{svjour}
\usepackage{graphicx,subfigure}
\usepackage{amsmath}
\usepackage[utf8]{inputenc}
\usepackage[T1]{fontenc}
\usepackage[english]{babel} 
\usepackage{mathtools}
\usepackage{wasysym}
\usepackage{bm}
\usepackage{float}
\usepackage{amssymb}
\usepackage{txfonts}
\usepackage{epstopdf}
\usepackage{dcolumn}
\usepackage[nottoc]{tocbibind}
\usepackage{wrapfig}
\usepackage{color}
\begin{document}
\title{Gravitational wave generation by interaction of high power lasers with matter using shock waves}
\author{Hedvika Kadlecov\'{a}\inst{1}\thanks{e-mail: hedvika.kadlecova@eli-beams.eu} \and Ond\v{r}ej Klimo\inst{1,2} \and Stefan Weber\inst{1} \and Georg Korn\inst{1}
}                     
%
%
\institute{Institute of Physics of the ASCR, ELI--Beamlines project, Na Slovance 2, 18221, Prague, Czech Republic \and FNSPE, Czech Technical University in Prague, 11519 Prague, Czech Republic}
\date{Received: date / Revised version: date}
%
\abstract{Gravitational wave generation by a strong shock wave in the interaction of high power laser with matter is analyzed in linear approximation of gravitational theory. The analytical formulas and estimates are derived for the metric perturbations and the radiated power of the generated gravitational waves. Furthermore the characteristics of polarization and the behavior of test particles are  investigated in the presence of gravitational wave which will be important for the detection. 
\PACS{{52.38.-r}{} \and {04.30.Db}{} \and {52.27.Ey}{} \and {52.38.Kd}{}} 
\keywords{gravitational waves -- laser -- plasma interaction -- generation of gravitational waves}
} 
\titlerunning{Gravitational wave generation by interaction of high power lasers}
\authorrunning{Hedvika Kadlecov\'{a} et al.}
\maketitle
\section{\label{sec:number1}Introduction}
The direct detection of gravitational waves remained one of the biggest challenges of experimental physics since the original paper by Einstein in 1918 \cite{Einstein1918} till February 2016 when interferometer measurements on LIGO and VIRGO confirmed the prediction \cite{AbbottGrav}. Their existence was indirectly shown thanks to the analytical work on radio pulses by Taylor and Hulse in 1974 who first recognized the pulsar PSR 1913+16 in a binary system \cite{Hulse1975}. The discovered pulsar was a unique binary pulsar which served as a perfect astrophysical laboratory for observations of very strong relativistic effects \cite{MTWBook}. 

The motivation for today's expensive experiments is the understanding of the universe and strong astrophysical sources because the gravitational waves carry the properties of the original source. By their detection we can obtain more information about the stars and the whole universe.
The information is hidden in the direction, amplitude, frequency and polarization of gravitational waves. The recent detection will definitely open a new era of experimental physics and astronomy.

The greatest obstacle in the detection of gravitational waves is that their intensity is very weak compared to electromagnetic waves. The gravitational force is the weakest one in the universe but affects the mass effectively on large distances and dominates the physics processes in the universe. 


Recent interests in astrophysical sources of high frequency gravitational waves (HFGW) with frequencies $\nu > 100$ kHz, GHz and higher lead to consider and revise the so called GW Hertz experiment which consists of generation and detection of the GW signal in terrestrial laboratories. 

The Hertz experiments in the high frequency domain were investigated by Chapline \cite{Chapline1974} and 
Rudenko \cite{Rudenko2004}. Namely, Rudenko suggested an experiment associated with high power electromagnetic waves and accoustic impulsive or shock waves travelling and interacting with a non--linear opto--acoustic media. Since lasers are the most powerful sources of electromagnetic radiation on Earth one may conceive schemes where gravitational waves are produced by interaction of laser with mediums in different setups  as was suggested in \cite{grossmannMeet2009l,izestELINP2014}. 

The gravitational waves are produced in space, for example, by non--symmetrical collapse of a star or by two black holes rotating tightly around each other. Such radiation is produced by a quadrupole moment -- the lowest multipole, whose change generates gravitational waves. The GW intensity is proportional to the time derivatives of the quadrupole moment up to third order, see Eqs.~(\ref{eq:QuadrupF}) and (\ref{eq:Luminosity}).

When an intense laser pulse interacts with a foil target, the target is ionized at the surface and free electrons are heated to a very high temperatures. Their thermal expansion from the target surface is accompanied by acceleration of mass inward due to momentum conservation and a shock wave eventually builds up. Such a process is non–symmetrical as the mass expanding outward is of a low density whereas the mass contained in the shock is compressed to a high density. Also the quadrupole moment of plasma changes in time thanks to increasing mass in the shock wave and possibly increasing its velocity as the interaction proceeds. Therefore with the advent of high energy lasers, the process of generating such waves starts to be an interesting scientific problem.

The first attempts to detect gravitational waves from space were reported by Weber in the low frequency domain \cite{Weber1959}. The Weber experiment constructed in 1960s consisted of a tube resonant detector. The detectors sensitivity was about $h \approx 10^{-16}$ which is still smaller than the amplitude of gravitational waves coming from space on Earth which is about $~10^{-18}$. The measurements made by Weber were not reproduced and it is assumed that some systematic mistake of the experiment was measured instead of gravitational waves. The work was followed by many groups. Today, the Weber type of detectors have sensitivities of $< 10^{-19}$ for example at Lousiana university where they use two-mode superconducting transducer and amplifier (SQUID) in the frequency range  $< 100$ Hz. Other examples are the projects MiniGrail \cite{MiniGrail} and Auriga in Padova (Italy)  \cite{Auriga,Auriga2}.

The gravitational waves (GW) generated from astrophysical sources are searched for by large gravitational interferometer detectors such as the American LIGO \cite{LIGO,LIGO2} and Italian--French VIRGO \cite{VIRGO}, for a review see \cite{Braginskij2000}. These detectors look for waves in the low frequency spectral band between 10 Hz and 20 kHz. The sensitivity of the LIGO type interferometers is about h =$10^{-23}$ which would allow to detect weak gravitational waves coming from  supernovea in our galaxy or Magellan Cloud. The experiments under construction are space-based interferometers, such as LISA \cite{LISA} and DECIGO \cite{DECIGO}. The vacuum environment would enable the interferometers get rid from the on Earth based noise. LISA would operate with $h > 10^{-22}$ and in a low frequency mode in the mHz range. The launch of L3 ESA program is planned for 2032. Other experiments in this area are GEO600 \cite{GEO} and CEGO (China) \cite{CEGO}.

Gravitational waves can be detected by many other techniques. They can be measured by change of lengths by extremely sensitive interferometers, piezoelectric crystals \cite{Baker2012}, superconductors, resonance chambers and by conversion of gravitational waves into electromagnetic waves by the Gertsenshtein effect \cite{LiFangYu2013,FangyuLi2009} or by sensors \cite{Arvanitaki2013}.

The main purpose of this paper is to analyze generation of high frequency gravitational waves (HFGW) in the interaction of high power laser pulse with matter. The scheme was suggested in \cite{grossmannMeet2009l,izestELINP2014} by using the material ablation \cite{Fabbro1984} and the radiation pressure \cite{Naumova2009}. The piston and light-sail models of ion acceleration have been analyzed in the context of gravitational waves in \cite{Gelfer2015}. We investigate the polarization of gravitational waves and the behavior of test particles in the gravitational field.

We suggest the generator type of gravitational wave experiment. The expected magnitude of wave perturbations is about $10^{-40}$ which is out of range of today's detectors. The technology of detectors as interferometers or resonant chambers for low frequency gravitational waves is useless for the high frequency range and a different technology is required. The detection problem of the generated GW is not addressed in this paper. The feasibility of detection of waves in high frequency regime is discussed for example in \cite{baker2005} for X-ray lasers, their detection in microwave band was discussed in \cite{FangyuLi2009}, detection by optically-levitated sensors with slightly higher frequency range than advanced LIGO was considered in \cite{LiFangYu2013}.

In 2011, a new detection scheme was proposed, called Li--Baker detector of HFGW \cite{LiBaker2011}, which might have sensitivities $~10^{-32}$ at $10$ GHz while the minimal detectable perturbation reaches $10^{-37}$. The detector is based on coupling between electromagnetic waves and GW in generalized inverse Gertsenshtein effect \cite{Gertsenshtein1962}. The scheme has also many opponents,\cite{JASON}, and is considered as non--realistic. Another technology should be developed to detect HFGW in the future. 

The remainder of the paper is organized as follows. In Section~\ref{sec:number2} we review the basic theory of linearized gravity which is used throughout the paper. We present the notation, the metric, the limitations of the theory, the analytical expressions for perturbations and luminosity, the polarization of gravitational radiation and the test particles behavior in the slow motion and distant field approximation of the linearized gravitational theory.  

In Section~\ref{sec:number11}, we review the shock wave model of the experiment for the gravitational waves generation under laboratory conditions. 
We derive the analytical formulas for the space perturbations and the luminosity of the gravitational radiation and give estimates for an experiment with megajoule class lasers.

In Section~\ref{sec:number5} we derive and analyze the polarization properties of gravitational radiation and the dependence on the orientation of the wave vector in the shock wave model.

Section \ref{sec:number6} concentrates on the GW detection with the analysis of the behavior of test particles in the field of gravitational waves.
 
The main results are summarized in Section~\ref{sec:number7}.

\section{\label{sec:number2}Linearized gravitational theory for gravitational waves studies}
In this section, we review the basics of linearized theory of gravitation \cite{MTWBook,MaggioreBook,BicakBook} needed for the subsequent analysis. We will use it to set up the notation which will be analyzed throughout the paper according to \cite{MaggioreBook}.
\subsection{Basics of the theory} 
The linearized theory of gravitation assumes the existence of a coordinate system in spacetime where the metric is close to the Minkowski flat metric $\eta_{\mu\nu}$ as
\begin{equation}
g_{\mu\nu}=\eta_{\mu\nu}+h_{\mu\nu},\quad \left|h_{\mu\nu}\right| \ll 1,
\label{eq:metric}
\end{equation}
where the perturbation of the metric is denoted as $h_{\mu\nu}$ and the 
Einstein equations read
\begin{equation}
\Box \bar{h}_{\mu\nu}=-\frac{16\pi G}{c^4}T_{\mu\nu}, \quad {\partial}^{\scriptscriptstyle \nu}\bar{h}_{\mu\nu}=0,\quad \bar{h}_{\mu\nu}=h_{\mu\nu}-\frac{1}{2}\eta_{\mu\nu}h,\label{eq:FinEinstein}
\end{equation}
where $\Box =\frac{1}{c^2}\frac{\partial^2}{\partial{t^2}}-\Delta$ is the d'Alembert operator, $T_{\mu\nu}$ is the stress--energy tensor, $G$ is gravitational constant and $c$ is the velocity of light. We have used the Lorentz gauge which reduces ten independent components of the $4\times 4$ matrix $h_{\mu\nu}$ to six independent components. Equations~(\ref{eq:FinEinstein}) satisfy the conservation of energy--momentum for consistency, $\partial^{\nu} T_{\mu\nu}=0$.

\subsection{Plane wave solution, TT gauge and polarization}
We investigate the propagation of gravitational waves and their interaction with test particles (and therefore with their detector) therefore we are interested in the solutions of (\ref{eq:FinEinstein}) outside of the sources $
\Box \bar{h}_{\mu\nu}=0,\, T_{\mu\nu}=0$. 
We consider the simplest solution for the wave equation (\ref{eq:FinEinstein}) which is the plane wave,
\begin{equation}
\bar{h}_{\mu\nu}= Re \left[A_{\mu\nu} \exp (ik_{\alpha}x^{\alpha})\right], \label{eq:PlaneW}
\end{equation}
where the $A_{\mu\nu}$ amplitude and $k_{\alpha}$ wave vector satisfy
$k_{\alpha}k^{\alpha}=0$. The $k_{\alpha}$ is a null vector and $A_{\mu\nu}k^{\alpha}=0$ so that $A_{\mu\nu}$ is orthogonal to $k_{\alpha}$. The solution describes a wave with the frequency $\omega/c\equiv k^{0}=(k_{x}^2+k_{y}^2+k_{z}^2)^{1/2}$ which propagates with the speed of light in the direction $(1/k^{0}(k_{x},k_{y},k_{z}))$.
The spatial components $\bar{h}^{TT}_{ij}$ of perturbation metric $h^{TT}_{\mu\nu}$ in the TT gauge\footnote{We use Latin letters ($i,j,\dots$) for transverse-space (spatial) indices and the Greek letters ($\mu,\nu$) for spacetime indices.} are obtained as
 \begin{equation}
\bar{h}^{TT}_{ij}=\Lambda_{ij,kl}h_{kl}, \label{eq:PlaneTT}
\end{equation}
where 
\begin{align}
\Lambda_{ij,kl}(\bm{n})=&P_{ik}P_{jl}-\frac{1}{2}P_{ij}P_{kl}=\delta_{ik}\delta_{jl}-\frac{1}{2}\delta_{ij}\delta_{kl}-n_{j}n_{l}\delta_{ik}-n_{i}n_{k}\delta_{jl}\nonumber\\
+&\frac{1}{2}n_{k}n_{l}\delta_{ij}+\frac{1}{2}n_{i}n_{j}\delta_{kl}+\frac{1}{2}n_{i}n_{j}n_{k}n_{l}. \label{eq:Projector}
\end{align}
The projector operator  is $P_{jk}(\bm{n})=\delta_{jk}-n_{j}n_{k}$ and $n_{k}=k_{k}/{k}$ is the unit vector in the direction of propagation. Note that this projector method is valid only for plane waves, \cite{MTWBook}. 

\subsection{Weak field sources with arbitrary velocity}
Equations~(\ref{eq:FinEinstein}) can be solved using Green functions and the leading term reads at the large distances as
\begin{equation}
h^{TT}_{ij}(t,\bm{x})=\frac{4G}{rc^4}\Lambda_{ij,kl}(\bm{n})\int d^{4}{x'}T_{kl}\left(t-\frac{r}{c}+\frac{\bm{x'}\cdot \bm{n}}{c},\bm{x'}\right). \label{eq:LargeH}
\end{equation}
Our convention for four--vector $k^{\mu}=(\omega/c,\bm{k})$ and $x^{\mu}=(ct,\bm{x})$ and then $k_{\mu} x^{\mu}=-\omega t + \bm{k}\cdot {\bm{x}}$ and $r=|\bm{x}-\bm{x}'|$.

\subsection{Low velocity multipole expansion}
In the non--relativistic system, ($v \ll c$) the wavelength of the radiation is much bigger than the size of the system,
\begin{equation}
\lambdabar \gg l_{d}, \label{eq:SlowV}
\end{equation}
where the linear size of the source is denoted as $l_{d}$ and reduced wavelength $\lambdabar=\lambda/2\pi$, \cite{MaggioreBook}. If this condition is valid need not to know the internal motions of the source and the radiation is dominated by the lowest multipole moments.

The perturbation (\ref{eq:LargeH}) can be expressed in multipole expansion, see \cite{MTWBook,MaggioreBook,BicakBook}. The lowest moment for gravitational waves is the quadrupole moment, the dipole moment vanishes because of the energy and momentum conservation laws. 
In the following text we work in the quadrupole approximation, which is the lowest term in the moment expansion (\ref{eq:LargeH}).

\subsection{Mass and quadrupole moment, luminosity}
The leading term in the expansion (\ref{eq:LargeH}) is 
\begin{align}
[h^{TT}_{ij}(t,\bm{x})]_{\text{quad}}=&\frac{1}{r}\frac{2G}{c^4}\Lambda_{ij,kl}(\bm{n})\ddot{I}^{kl}(t-r/c), \label{eq:QuadrupF}
\end{align}
where $I^{ij}$ denotes the quadrupole moment defined as 
\begin{align}
I^{ij}=M^{ij}-\frac{1}{3}\delta^{ij}M_{kk}
      \equiv \int d^{3}x\rho(x^{i}x^{j}-\frac{1}{3}r^2\delta^{ij}),\label{eq:QuadrupM}
\end{align}
and $\rho=\frac{1}{c^2}T^{00}$ becomes the mass density (in the lowest order in $v/c$). The mass moment is defined as
\begin{equation}
M^{ij}=\int d^{3}x\rho(t,\bm{x})x^{i}x^{j}. \label{eq:massMoment}
\end{equation}

The total gravitational luminosity (power) of the source in quadrupole approximation is
\begin{align}
\mathcal{L}_{\text{quad}}=&\frac{G}{5c^5}\langle \dddot{I}_{ij}\dddot{I}_{ij}\rangle, \label{eq:Luminosity}
\end{align}
where $I_{ij}$ is evaluated in the retarded time $t-r/c$ if it is not specified. 

In the following text, we consider the entities $I_{ij}$ and $h^{TT}_{ij}$ evaluated in the retarded time $t-r/c$ if it is not specified explicitly.


\section{\label{sec:number11} Derivation of gravitational wave characteristics for the shock wave model}
This section is devoted to the derivation of fully analytical formulae of the luminosity ${\mathcal{L}}_{GW}$ and the perturbation of the metric $h_{GW}$ for the shock wave model using the linearized gravity theory from Section~\ref{sec:number2}. 
 
 
\subsection{Shock wave model}
The shock wave model is described according to the analytical theory of planar laser-driven ablation \cite{Fabbro1984}.
The process is a function of the laser intensity, wavelength, the target material, and the degree of inhibition of the electron thermal conduction. 

In this configuration, the laser is interacting with a planar thick foil of about 1 mm thickness. This thickness corresponds to the distance the shock wave can travel inside the target for the laser and target parameters given in Section 3.3.4. The material is accelerated in the ablation zone and in the shock front. 
Here, we limit our analysis to the shock wave acceleration where the target density is higher and it makes larger contribution to the GW generation.


In the shock wave model, the laser launches a shock with the velocity $v_{s}$. The material from the foil is accelerated along the axis of motion $z$ and it produces a change of quadrupole mass moment of the source $I_{ij}$. A typical order of the laser pulse duration is $1$ ns therefore the gravitational waves are generated in the GHz domain.


\subsection{\label{sub:limit}Limitations of the theory}
In our non--relativistic model the wavelength of the generated radiation must be much bigger than the linear size of the source (\ref{eq:SlowV}), 
This is valid as the linear size of the source is $l_{d} < 1\,$mm (target thickness).

The expected GW wavelength is 30 cm, which is larger than the source size.  
Therefore the low velocity condition (\ref{eq:SlowV}) is satisfied for our shock wave experiment. 
In general, the limitation (\ref{eq:SlowV}) in the laboratory conditions is much stringent than for a pulsar in space, where the difference is 15 orders of magnitude. 

\subsection{Shock wave model calculations}
We choose the orthogonal coordinate system $x, y, z$ and we assume the whole process takes place in a box of rectangular shape with parameters $a, b, l$ for simplicity. The origin of the coordinate system corresponds to the position of the foil front surface.


The moving shock front where the density of the beam changes is denoted as $z_{s}$ with the functional dependence
\begin{align}
z_{s}(t)= v_{s}t, \label{eq:zss}
\end{align}
where the shock front velocity is defined as
\begin{align}
v_{s}\simeq \sqrt{\frac{P_{s}}{\rho_{0}}}, \label{eq:vs}
\end{align}
where $P_{s}$ is shock wave pressure\footnote{In the case of subsonic deflagration assumed in this paper, the difference between ablation and shock pressure is neglected, because depending on the energy deposition at the heat front is at most 2 in the stationary ablation model \cite{AtzeniBook}.}, $\rho_{0}$ is material density and $f_{1}$ is the distance from the shock wave front to the detector.

We assume that for $t=0$, $z_{s}(0)=0$, which is our convenient choice. 
We use general expression for $P_{s}$, which allows us to have more control parameters than the one used in \cite{grossmannMeet2009l,izestELINP2014} for a fixed wavelength. The relation of $P_{s}$ and $I_{L}$ (laser intensity) is the following, \cite{AtzeniBook},
\begin{equation}
P_{s}=(\rho_{c}I^{2}_{L})^{1/3},\; \rho_{c}=\frac{\epsilon_{0}m_{e}m_{i}}{Ze^2}\frac{(2\pi c)^2}{\lambda_{L}^2},\label{eq:Pressure}
\end{equation}
where $\rho_c$ is the critical mass density, $\epsilon_{0}$ is permitivity of vacuum, $m_{e}$ is the rest mass of the electron, $m_{i}$ is the mass of the ion, $e$ is the charge of electron, $\lambda_{L}$ is the wavelength of the laser, $Z$ is the atomic number. All of the parameters in $\rho_{c}$ are constants except the laser wavelength $\lambda_{L}$ which is given in the specific experiment. 


In the following, we calculate everything with a general function $z_{s}(t)$ and then substitute the explicit function (\ref{eq:zss}) at convenient places. General expressions might be useful for other forms of $z_{s}(t)$. 
The shock wave density reads 
\begin{equation}
\rho(\bf{x})=
\begin{cases}
\;4\rho_{0}  &\quad if \quad \tfrac{3}{4}z_{s}<z<z_{s},\\
\;\rho_{0} &\quad if \quad z_{s}<z<l,\label{eq:densityShock1}\\
\;0 &\quad \quad \text{otherwise},
\end{cases}
\end{equation}
it satisfies the mass conservation condition.

\subsubsection{Mass moment}
The first step is the mass moment derivation.
The transverse plain is covered by two spatial coordinates $x^{i}=\{x,y\}$, the transverse space ${\rm d}{s^2_{\perp}}=g_{ij\perp}{\rm d}x^{i}{\rm d}x^{j}={\rm d}x^2+{\rm d}y^2$ is flat and is characterized by a vector $\zeta^{i}=\{a,b\}$.
The mass moment (\ref{eq:massMoment}) is a $3 \times 3$ symmetric matrix in spacelike coordinates, therefore it is necessary to calculate just the components on diagonal $M_{ii}, M_{zz}$ and non--diagonal components $M_{ij}, M_{iz}$:

\begin{align}
M_{ii}&=\frac{S\rho_{0}}{3}\zeta_{i}^2l,\; M_{zz}=\frac{S}{3}\rho_{0}\left(\frac{21}{16}z^{3}_{s}+l^3\right),\nonumber\\
M_{ij}&=\frac{S^2\rho_{0}}{4}l,\, 
M_{iz}=\frac{S\rho_{0}}{4}\zeta_{i}\left(\frac{3}{4}z^2_{s}+l^2\right)\label{eq:massequiv},
\end{align}
where $S=ab$ is the shock front surface.

\subsubsection{Quadrupole moment}
The next step is the calculation of the quadrupole moment (\ref{eq:QuadrupM}). The second term is non-zero only for the diagonal components. The non--diagonal components $I_{ij}, I_{iz}$ are
\begin{align}
I_{ij}&=M_{ij},\quad I_{iz}=M_{iz}.\label{eq:componentsQuadrMomentOff}
\end{align}
The diagonal components $I_{ii}=M_{ii}-\frac{1}{3}Tr M$ read
\begin{align}
I_{ii}&=\frac{S\rho_{0}}{9}\left\{ (2\zeta_{i}^2-\zeta_{j}^2-l^2)l-\frac{21}{16}z^3_{s}\right\},\nonumber\\
I_{zz}&=\frac{S\rho_{0}}{9}\left\{ (2 l^2-\zeta_{i}^2-\zeta_{j}^2)l+\frac{21}{8}z^3_{s}\right\},\nonumber\\.\label{eq:componentsQuadMomentDiag}
\end{align}
The diagonal components of quadrupole moment show cubic dependence on the function $z_{s}$ and are missing a quadratic term. The non-diagonal components $I_{iz}$ are missing the linear dependence on $z_{s}$.

\subsubsection{Analytical form of perturbation and luminosity}
Now, we calculate the components of the perturbation tensor according to (\ref{eq:QuadrupF}) without projector appearing in (\ref{eq:PlaneTT}) in a general form:
\begin{align}
h_{ij}=0,\quad
h_{zz}=\frac{7}{12}\frac{GS\rho_{0}}{rc^4}\frac{\partial^2 z^3_{s}}{\partial t^2},\label{eq:DiagHpert}
\end{align}
and the non-diagonal terms are
\begin{align}
h_{ii}=-\frac{7}{24}\frac{GS\rho_{0}}{rc^4}\frac{\partial^2 z^3_{s}}{\partial t^2},\quad h_{iz}=\frac{3}{8}\frac{GS\rho_{0}}{rc^4}\zeta_{i}\frac{\partial^2 z^2_{s}}{\partial t^2}.\label{eq:NodiagHpert}
\end{align}



After substituting the quadrupole moment components into (\ref{eq:Luminosity}), we get the general expression
\begin{align}
\mathcal{L}_{\text{quad}}&=\frac{GS^2\rho^2_{0}}{320c^5}\left[\frac{49}{6}\left(\frac{\partial^3 z^3_{s}}{\partial t^3}\right)^2+\frac{9}{2}(\zeta^2_{i}+\zeta^2_{j})\left(\frac{\partial^3 z^2_{s}}{\partial t^3}\right)^2\right] \label{eq:LumRes2}
\end{align}
The explicit substitution $z_{s}$ simplifies the expression (\ref{eq:LumRes2}) that just the diagonal components of quadrupole moment contribute to the result, which reads
\begin{align}
\mathcal{L}_{\text{quad}}=&\frac{147}{160}\frac{GS^2\rho_{0}^2v^6_{s}}{c^5}=\frac{147}{160}\frac{G P^2_{L}}{c^5}\frac{\rho_{c}}{\rho_{0}}. \label{eq:LumResFin}
\end{align}
and further using (\ref{eq:vs}) and (\ref{eq:Pressure}), we will obtain the final expression for luminosity of gravitational radiation. The luminosity then depends on the laser power $P_{L}=I_L S$, where $I_L$ is the laser intensity, the density of the material and the laser wavelength. The numerical factor in Eq.~(\ref{eq:LumResFin}) is presented in Section~\ref{subsec:estimate}.

The perturbation of the space $h^{GW}_{zz}$ using 
(\ref{eq:vs}), (\ref{eq:Pressure}) and (\ref{eq:QuadrupF}) reads
\begin{align}
h_{zz}&=\frac{7}{2}\frac{GE_{L}}{rc^4}\sqrt{\frac{\rho_{c}}{\rho_{0}}},\label{eq:DiagHpv}
\end{align}
where $E_{L}=SI_{L}\tau$ is the laser energy and $\tau$ is the laser pulse duration. We have assumed stationary shock wave where the dissipated energy is recovered from the absorpted laser energy, and the time of GW emission is equal to the laser pulse duration. 
 
This formula for the perturbation of the space by gravitational wave 
generalizes the result obtained in \cite{grossmannMeet2009l,izestELINP2014}, to an arbitrary laser wavelength. 

The value of perturbation decreases with the distance as $1/r$. The numerical factors are evaluated in the next section for specific experimental parameters.



\subsubsection{\label{subsec:estimate}Numerical estimates}

The values for luminosity (\ref{eq:LumResFin}) and the perturbation $h^{GW}_{zz}$ of the space by the gravitatinal wave in $zz$ direction, (\ref{eq:DiagHpv}), for realistic experimental parameters of the megajoule scale lasers \cite{NIF,LMJ}:
\begin{align}
\mathcal{L}_{\text{quad}}[{\rm W}]=2.53\times 10^{-23}\frac{\rho_{c}}{\rho_{0}}P^2_{L}[\rm PW]. \label{eq:NumericalLuminosity}
\end{align}
Our results are in agreement with \cite{grossmannMeet2009l,izestELINP2014}. 
We estimate the $zz$ component of the perturbation tensor $h_{ij}$ (\ref{eq:DiagHpv}), 
similarly to the previous case, we obtain
\begin{align}
h_{zz}=&2.89\times 10^{-38}
\sqrt{\frac{\rho_{c}}{\rho_{0}}}\frac{E_{L}[\rm MJ]}{r[\rm m]}.\label{eq:NumericalHzz}
\end{align}
 
The value of the critical mass density is estimated for a Carbon target with $A=12,\,Z=6$, wavelength $\lambda_{L}=351$ nm, $\rho_{c}=15\,{\rm mg}/{\rm cm}^3$.
For the laser parameters we consider the megajoule scale installations NIF \cite{NIF} and LMJ \cite{LMJ}, $P_{L}=0.5 \text{PW}$, $E_{L}=0.5\,\text{MJ}$, $\tau =1{\text{ns}}$, the detection distance $r=10$ m and  the target foil of the density $\rho_{0}=30\,\text{mg}/\text{cm}^3$ and the size $a=b=l=1$ mm.


The reason for this choice is that the metric distortion as given in (\ref{eq:DiagHpv}) is directly proportional to laser pulse energy and thus the megajoule class lasers can maximize emission of gravitational waves. The third-harmonic of the fundamental frequency of these lasers (corresponding to $\lambda_{L}=351$~nm) makes the absorption of the laser light more efficient. 

The outgoing gravitational radiation has a pulse duration of $1$ ns which is given by the laser pulse length. The estimations for our expressions of the luminosity and the perturbation are:
\begin{align}
{\mathcal{L}}_{GW}&\simeq 7.86\times 10^{-25}\,{\rm W},\quad h^{GW}_{zz}\simeq 5.1\times 10^{-40}.\label{eq:FinalResults}
\end{align}
These estimates agree with those presented in \cite{grossmannMeet2009l,izestELINP2014}.



\section{\label{sec:number5}Polarization of gravitational waves}
In this section to investigate two linearly independent polarization modes of the gravitational waves, $+$ and $-$, and focus on their interpretation which would be useful for planning the experiments.

We can decompose a gravitational wave into two linearly polarized components or into two circularly polarized components as
\begin{align}
h^{TT}_{ij}({\bm x})&=h_{+\,ij}+h_{\times\,ij},\label{eq:poLinHH}
\end{align}
where we introduce unit linear tensors of linear polarization as $
e_{+\,ij}=({\bf e}_{1})_{i}({\bf e}_{1})_{j}-({\bf e}_{2})_{j}({\bf e}_{2})_{i},\,
e_{\times\,ij}=({\bf e}_{1})_{i}({\bf e}_{2})_{j}+({\bf e}_{1})_{j}({\bf e}_{2})_{i}$, where we have denoted ${\bf e}_{1}$ and ${\bf e}_{2}$ as unit vectors.

The two independent linearly polarized waves can be written as
\begin{align}
h_{+ij}&=A_{+}e^{-i\omega(t-{\bf n}\cdot {\bf x})}e_{+ij},\quad
h_{\times ij}=A_{\times}e^{-i\omega(t-{\bf n}\cdot {\bf x})}e_{\times ij},\label{eq:hxcomp}
\end{align}
where $A_{+}$ and $A_{\times}$ are amplitudes of the two independent polarizations.
For example ${\bf e}_{1}=(1,0,0)$ and ${\bf e}_{2}=(0,1,0)$ for the wave vector ${\bf n}=(0,0,1)$\textcolor{red}{,} we will get the result for wave propagation in $z$ coordinate.


We can investigate the gravitational waves in dependence on the direction of the propagation, i.e. on the wave vector. The set up of our models is that the direction of the motion is the $z$ coordinate. Therefore we look at the general orientation of the wave vector in spherical coordinates and analyze the space distribution of the gravitational waves from the experiment. These results could help in specifying the exact position for detectors in experiment.

\subsection{Angular dependence of the wave amplitude}

The general direction of the wave propagation can be expressed in the spherical coordinates as
\begin{equation}
\bm{n} = (\sin\theta\sin\phi,\,\sin\theta\cos\phi,\,\cos\theta),
\label{eq:ngen}
\end{equation}
and the perturbation tensor can be obtained via (\ref{eq:QuadrupF}) and the projector (\ref{eq:Projector}).

The general expressions for the two modes of polarizations are, \cite{MaggioreBook},
\begin{align}
A_{+}(t;\theta,\phi)=&\frac{1}{r}\frac{G}{c^4}\left[\ddot{M}_{xx}(\cos^{2}\phi-\sin^2\phi\cos^{2}\theta)\right.\nonumber\\
&+\left.\ddot{M}_{yy}(\sin^2\phi-\cos^2\phi\cos^2\theta)-\ddot{M}_{zz}\sin^2\theta\right.\label{eq:hpn}\\
&-\left.\ddot{M}_{xy}\sin{2\phi(1+\cos^2\theta)}+\ddot{M}_{xz}\sin\phi\sin{2\theta}\right.\nonumber\\
&+\left.\ddot{M}_{yz}\cos\phi\sin{2\theta}\right], \label{eq:hkn}\\
A_{\times}(t;\theta,\phi)&=\frac{1}{r}\frac{G}{c^4}\left[(\ddot{M}_{xx}-\ddot{M}_{yy})\sin{2\phi}\cos\theta-2\ddot{M}_{xz}\cos\phi\sin\theta\right.\nonumber\\
&+\left.2\ddot{M}_{xy}\cos{2\phi}\cos\theta+2\ddot{M}_{yz}\sin\phi\sin{\theta}\right],
\end{align}
and the whole components of $h^{TT}_{ij}$ can be expressed as (\ref{eq:poLinHH}) and (\ref{eq:hxcomp}).
Afterwards we express mass moments in terms of derivatives of function $z_{s}$, the amplitudes read as follows,
\begin{align}
A_{+}(t;\theta,\phi)&=\frac{1}{16}\frac{GS\rho_{0}}{rc^4}\left[3\sin{2\theta}(a\sin\phi+b\cos\phi)\frac{\partial^2 z^2_{s}}{\partial t^2}\right.\label{eq:hpn1}\\
&\left.-7\sin^2\theta\frac{\partial^2 z^3_{s}}{\partial t^2}\right],\nonumber\\
A_{\times}(t;\theta,\phi)&=\frac{3}{8}\frac{GS\rho_{0}}{rc^4}\sin\theta(b\sin\phi-a\cos\phi)\frac{\partial^2 z^2_{s}}{\partial t^2}.\label{eq:hkn1}
\end{align}

After we use the ansatz for the $z_{s}$, we get
\begin{align}
A_{+}(t;\theta,\phi)&=B_{C}P_{A_{+}}(t;\theta,\phi), \quad
A_{\times}(t;\theta,\phi)=B_{C}P_{A_{\times}}(\theta,\phi),\label{eq:hkn2}\\
B_{C}&=\frac{3}{8}\frac{GS\rho_{0}}{rc^4}v^2_{s}=\frac{3}{8}\frac{GS\rho_{0}}{rc^4}(\rho_{c}I^2_{L})^{1/3},\label{eq:hk3}
\end{align}
where the angular dependence is denoted as
\begin{align}
P_{A_{+}}(t;\theta,\phi)&=\frac{1}{r}\big\{\sin{2\theta}(a\sin\phi+b\cos\phi)-7\sin^2 \theta v_{s}t\big\}, \nonumber\\
P_{A_{\times}}(\theta,\phi)&=\frac{1}{r}\sin\theta(b\sin\phi-a\cos\phi).\label{eq:PAtimes}
\end{align}
The amplitude $A_{+}$ increases linearly in time during the laser pulse and $A_{\times}$ is constant in time. The amplitudes decrease with the radial distance as $1/r$. 

\begin{figure}[h]
\centering
\includegraphics[width=0.45\textwidth]{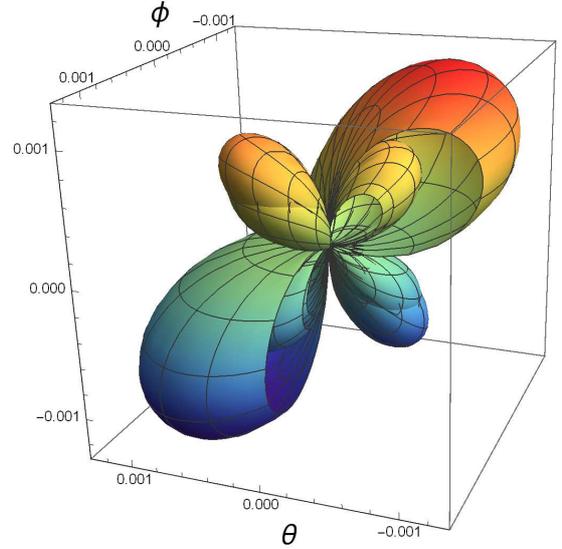}
\caption{\label{fig:Ampl+}(Color online) Dependence of the angular part of amplitude $P_{A_{+}}(t;\theta,\phi)$ (\ref{eq:PAtimes}) on the angles $\theta$ and $\phi$ at two distances from the source: the biggest surface corresponds to $r=1$ m and then $r=1.5$ m. The distance from the center is proportional to the wave amplitude. The squeezed toroid with cloverleaf shape is cut on purpose to see the inner surface of lower $r$. }
\end{figure}

In Fig.~\ref{fig:Ampl+} we demonstrate the effect of time dependence of the $A_{+}$ amplitude. The angular shape of $P_{A_{+}}(t;\theta,\phi)$ of the shock wave is depicted in Fig.~\ref{fig:Ampl+} for $l/v_{s}>t>0$. The angular dependence has a shape of a squeezed toroid with the center at the target position $z=0$ given by $\sin^2\theta$ term with cloverleaf shape which is a contribution from the term with $\phi$ angle. The surfaces inside of the angular shape represent angular structure for larger $r$ and we observe that the magnitude of the angular shape becomes smaller, as expected, as $1/r$.

The numerical application is made for the same parameters as in Section~\ref{subsec:estimate}: the target size $a=b=l=1$ mm, the critical density $15\,\text{mg}/\text{cm}^3$ and laser intensity $I_L=50\,\text{PW}/\text{cm}^2$ and the shock velocity $v_s=1.5\times 10^6$ m$/$s. 

\begin{figure}[h!]
\centering
\includegraphics[width=0.55\textwidth]{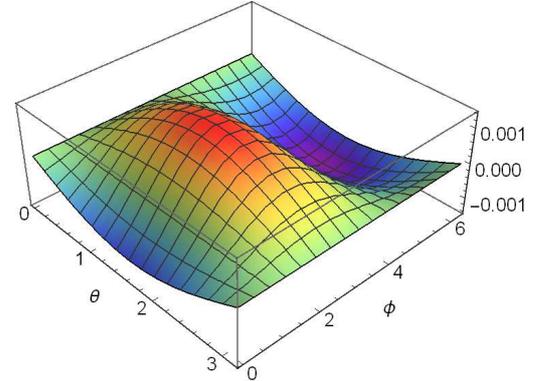}
\caption{\label{fig:AmplCross} (Color online) The angular part of amplitude $P_{A_{\times}}(\theta, \phi)$ (\ref{eq:PAtimes}) is pictured in dependence on $\theta$ and $\phi$ angle in radians at any time.}
\end{figure}

The amplitude for polarization mode $\times$ is shown in Fig.~{\ref{fig:AmplCross}}. It starts at $z=0$ that is, at the source, and its amplitude reaches its maximum and then minimum. 

There are three particular cases of interest. In the case of parallel propagation, the case $n_{z}=z$ for $\theta=0^{\circ},\,\phi=0^{\circ}$, both amplitudes equal to zero, in the case $n_{x}=x$ for $\theta=90^{\circ},\, \phi=90^{\circ}$ and case $n_{y}=y$ for $\theta=90^{\circ},\,\phi=0^{\circ}$ the waves' amplitude attain their maximum values.

The GW is transverse wave and it is not emitted in the direction of the shock propagation $n_{z}=z$. Lets look closely on the cases $n_{x}=x$ and $n_{y}=y$.

\subsubsection{\label{sub:x}Wave propagation in the $x$ and $y$--direction}
The $h^{TT}_{ij}$ (\ref{eq:PlaneTT}) has the only non--vanishing components for the wave vector in the $x$--direction $n=(1,0,0)$,
\begin{align}
h^{TT}_{yy}=-h^{TT}_{zz}=\text{Re}\left\{A^{x}_{+}e^{-i\omega(t-x/c)}\right\},\;
h^{TT}_{zy}=h^{TT}_{yz}=\text{Re}\left\{A^{x}_{\times}e^{-i\omega(t-x/c)}\right\},\label{eq:hTTcomponentsX}
\end{align}
where we have used the definition of perturbation tensor (\ref{eq:QuadrupF}) together with the only non--zero components of the projector $\Lambda_{ij,kl}$ (\ref{eq:Projector}).


The waves are linearly polarized. We obtain the amplitudes of the polarization modes from Eqs.~(\ref{eq:hpn1}), (\ref{eq:hk3}) and (\ref{eq:PAtimes}), where we set $\theta=90^{\circ},\, \phi=90^{\circ}$:
\begin{align}
A^{x}_{+}&=\frac{1}{r}\frac{G}{c^4}(\ddot{M}_{yy}-\ddot{M}_{zz})=-\frac{7}{16}\frac{GS\rho_{0}}{rc^4}\frac{\partial^2 z^3_{s}}{\partial t^2}, \nonumber\\
A^{x}_{\times}&=\frac{2}{r}\frac{G}{c^4}\ddot{M}_{yz}=\frac{3}{8}\frac{GS\rho_{0}b}{rc^4}\frac{\partial^2z^2_{s}}{\partial t^2}\label{eq:hkx}.
\end{align}

By using the expression for $z_{s}$, we get
\begin{align}
A^{x}_{+}&=-\frac{21}{8}\frac{GS\rho_{0}}{rc^4}{v^{3}_{s}}t=-\frac{21}{8}\frac{GE_{L}\rho_{0}}{rc^4}\sqrt{\frac{\rho_{c}}{\rho_{0}}},\nonumber \\
A^{x}_{\times}&=\frac{3}{4}\frac{GS\rho_{0}b}{rc^4}{v_{s}^{2}}=\frac{3}{4}\frac{GS\rho_{0}b}{rc^4}(\rho_{c}I^2_{L})^{1/3}\label{eq:hkx2}.
\end{align}


The perturbation tensor in $TT$ calibration (\ref{eq:PlaneTT})
for the wave vector in the $y$--direction $n=(0,1,0)$ reads
\begin{align}
h^{TT}_{xx}=-h^{TT}_{zz}=\text{Re}\left\{A^{y}_{+}e^{-i\omega(t-y/c)}\right\},\;
h^{TT}_{zx}=h^{TT}_{xz}=\text{Re}\left\{A^{y}_{\times}e^{-i\omega(t-y/c)}\right\},\label{eq:hTTcomponentsY}
\end{align}
then we obtain the amplitudes of the polarization modes from Eqs.~(\ref{eq:hpn1}), (\ref{eq:hk3}) and (\ref{eq:PAtimes}), where we set $\theta=90^{\circ},\,\phi=0^{\circ}$:


\begin{align}
A^{y}_{+}=\frac{1}{r}\frac{G}{c^4}(\ddot{M}_{xx}-\ddot{M}_{zz})=A^{x}_{+},\quad
A^{y}_{\times}=-\frac{2}{r}\frac{G}{c^4}\ddot{M}_{xz}=-A^{x}_{\times}\label{eq:hky}.
\end{align}

The two cases are symmetrical, the resulting amplitudes $A^{y}_{+}$ and $A^{y}_{-}$ have the same form as for the direction $x$ apart from the sign in $A^{y}_{\times}$.
The amplitudes are non--zero for both '$+$' and '$\times$' polarization modes. They depend on the focus area $S$, the density of the material $\rho_{0}$ and the velocity of the ions $v_{s}$. The amplitudes decrease with the radial distance like $1/r$.

\subsection{\label{fig:Spectrum}Angular dependence of the emitted energy}
The energy and momentum $t^{GW}_{\mu\nu}$ carried by the GWs at large distances from the source (at the position of the detector) is described by the effective tensor
\begin{align}
t^{GW}_{\mu\nu}=&\frac{c^4}{32\pi G}\langle \partial_{\mu}h_{\alpha\beta}\partial_{\nu}h^{\alpha\beta}\rangle,\quad t^{GW}_{00}=\frac{c^4}{32\pi G}\langle \dot{A}^2_{+}+ \dot{A}^2_{\times}\rangle\label{eq:StressEnergyTenz},
\end{align}
where the brackets mean averaging over the laser pulse duration, and the tensor (\ref{eq:StressEnergyTenz}) satisfies the energy conservation $\partial^{\mu}t_{\mu\nu}=0$, \cite{MaggioreBook,BicakBook}, and $t^{GW}_{00}$ denotes gauge invariant energy density.
Both amplitudes of modes  $+$ and $\times$ contribute to the gauge density (\ref{eq:StressEnergyTenz}) even though the amplitude $A_{\times}$ is time independent because they are evaluated in retarded time.
The energy spectrum is then $\frac{d E}{d A}=\frac{c^3}{16\pi G}\int_{0}^{\tau}d t (\dot{A}^2_{+}+\dot{A}^2_{\times})=\frac{c^3}{16\pi G}(\dot{A}^2_{+}+\dot{A}^2_{\times})\tau$ where $d A=r^2 d \Omega$ is surface element and $\tau$ is duration of pulse. The energy spectrum is defined by the laser pulse shape.

The radiative characteristic represents amount of energy going through various directions in the case of large distances $r$ for quadrupole radiation. We can obtain it by using (\ref{eq:StressEnergyTenz}) for the radial component of the vector of Poynting  as
\begin{align}
-t_{0r}^{GW}\equiv \mathcal{S}=\frac{c^3}{72G\pi r^2}&\langle\dddot{I}_{ij}\dddot{I}_{ij}-2\dddot{I}_{is}\dddot{I}_{sj}n_{i}n_{j}+\frac{1}{2}\dddot{I}_{ij}\dddot{I}_{rs}n_{i}n_{j}n_{r}n_{s}\rangle.
\label{eq:RadChar}
\end{align}

The luminosity ${\mathcal{L}}_{GW}$  defined in (\ref{eq:Luminosity}) as a radiated power, it is connected to $t_{\mu\nu}^{GW}$ as $-\frac{d E}{d t}=\mathcal{L}_{GW}=\int t_{0r}^{GV}r^2\sin\theta d \theta d \phi$,
which can be found by integration over a sphere with radius $r$.



We can evaluate (\ref{eq:RadChar}) for an arbitrary emission direction  as
 \begin{align}
\mathcal{S}_{n}=&\frac{c^3}{72G\pi r^2}\left[(\dddot{I}_{xx})^2(1-2\sin^{2}\theta\sin^{2}\phi)\right.\label{eq:RadCharn}\\
+&\left.(\dddot{I}_{yy})^2(1-2\sin^{2}\theta\cos^{2}\phi)+(\dddot{I}_{zz})^2(1-2\cos^{2}\theta)\right.\nonumber\\
+&\frac{1}{2}\left.(\dddot{I}_{xx}\sin^{2}\theta\sin^{2}\phi+\dddot{I}_{yy}\sin^{2}\theta\cos^{2}\phi+\dddot{I}_{zz}\cos^{2}\theta)^2\right].\nonumber
\end{align}

Using the expression for the shock wave position $z_s$, the expression for the Poynting vector reads
and the expression for the general wave vector (\ref{eq:ngen}) results in
\begin{align}
\mathcal{S}_{n}=B_{S_n}P_{S_{n}}(\theta,r),\quad B_{S_n}=&\frac{49}{9216}\frac{S^2c^3{\rho_{0}^2}{v_{s}^{6}}}{G\pi},\label{eq:radCharSn}
\end{align}
where the angular dependence is given as
\begin{align}
P_{S_n}(\theta,r)= \frac{1}{r^2}\left[12-4(4\cos^2{\theta}+\sin^{2}\theta)+(2\cos^{2}\theta-\sin^{2}\theta)^2\right].\label{fig:P}
\end{align}

The radiative characteristics (\ref{fig:P}) depend only on the $\theta$ angle, the angle $\phi$ cancels out during the derivation because $\dddot{I}_{xx}=\dddot{I}_{yy}$. It behaves as $1/r^2$ as expected for a spherical wave.

\begin{figure}[h!]
\centering
\includegraphics[width=0.45\textwidth]{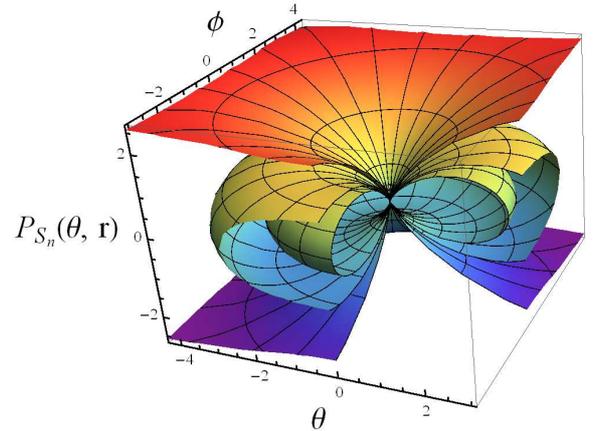}
\caption{\label{fig:CharRad} (Color online) The radiation characteristics $\mathcal{S}_{n}$ (\ref{eq:radCharSn}) is pictured in dependence on $\theta$ angle and rotated additionally around $\phi$ angle in radians. We have plotted just the angular dependence $P_{S_{n}}(\theta,r)$ (\ref{fig:P}), $\mathcal{S}_{n}=B_{S_n}P_{S_{n}}(\theta,r)$, (\ref{eq:radCharSn}). The dependence on $1/r^2$ is depicted in smaller surfaces in the figure, the biggest surface is $r=1$ m, then $r=1.5$ m, $1.8$ m. The surface is getting smaller as $r\rightarrow 10$ m (the distance of the detector). 
The structure of surfaces is symmetric around the axes $z=0$. The structure of the surface has similar shape as the amplitudes.}
\end{figure}

The angular dependence of the Poynting vector is shown in Fig.~\ref{fig:CharRad}, where the structure of the surface has a toriodal shape similar to the amplitudes (\ref{eq:PAtimes}).

 
This directional analysis confirms the preferential wave emission in the transverse plane with maxima in the diagonal directions of the square target.

\section{\label{sec:number6}Behavior of test particles in the presence of a gravitational wave}
In this section we investigate the influence of the gravitational wave on two closely placed particles 
by analysing the distance between them. 
The equation of deviation is defined as
\begin{equation}
\frac{d^2 x^j_{B}}{dt^2}=-R^{TT}_{j0k0}x^{k}_{B}=\frac{1}{2}\frac{\partial^2 h^{TT}_{jk}}{\partial_{t}^2}x^{k}_{B},\label{eq:deviation}
\end{equation}
which describes the relative acceleration of two particles originally moving along two parallel trajectories in proper detector frame. We have used the fact that to first order in $h^{TT}_{ij}$, $t=\tau + O (h)$. The equation (\ref{eq:deviation}) can be integrated  if we assume that the particles are at rest relative to each other before the wave arrives (when $h_{ij}$=0 then $x_{B}^{j}=x_{B(0)}^j$). The equation of motion yields $
x_{B}^j(\tau)=x_{B(0)}^{k}\left[\delta_{jk}+\frac{1}{2}h^{TT}_{jk}(\tau)\right]_{\text{at position of A}}$
and describes the oscillations of particle B measured in the reference frame of particle A. The position of the second particle B can be affected by the wave just in traversal directions to the wave propagation direction.


\subsection{Conditions on the detector}
The equation of deviation (\ref{eq:deviation}) is valid as long as $|x_{B}|$ is much smaller than typical scale over which the gravitational field changes substantially. We consider the laser driven GWs in the GHz domain as a monochromatic wave with the characteristic wavelength of $30$ cm and the detector size is supposed to be much smaller than that. For example, the bar detectors and ground based interferometers may satisfy this condition but LISA does not.

\subsection{Movement of particles}
First, we investigate the effect of the mode $+$ for the wave vector in $x$ direction (\ref{sub:x}). Since the shock is propagating in the $z$ direction and the wave vector is oriented along the $x$ coordinate, the particle is moving in the $(z,y)$ plane. We assume that the test particles are distributed on a circle of radius $r_{c}$ at time $\tau=0$. In the center of the circle is the reference particle A, which is at rest in the proper reference frame. The position of the particle on the circle in any time is defined by (\ref{eq:deviation}). We denote the coordinates in the reference frame $\tilde{x},\, \tilde{y}$ and $\tilde{z}$, as
\begin{align}
\tilde{z}_{B}(\tau)&=\left[1+\frac{1}{2}h^{TT}_{zz}(\tau)|_{\tilde{x}_{A}^{j}=0}\right]\tilde{z}_{B(0)},\nonumber\\
\tilde{y}_{B}(\tau)&=\left[1-\frac{1}{2}h^{TT}_{zz}(\tau)|_{\tilde{x}_{A}^{j}=0}\right]\tilde{y}_{B(0)}, \label{eq: devCalc}
\end{align}
where we have used $h^{TT}_{zz}=-h^{TT}_{yy}$, $z_{B(0)}=r_{c}\cos\chi$ and $y_{B(0)}=r_{c}\sin \chi$ which are coordinates in time $\tau=0$ when the particles were distributed along the circle of radius along the azimuthal angle $\chi$. In the further text, we assume all $h_{ij}^{TT}|_{\tilde{x}_{A}^{j}=0}$ (i.e. are evaluated at $\tilde{x}_{A}^{j}=0$) and we will not write it explicitly.
 
Expression (\ref{eq: devCalc}) describes an ellipse with semi-minor axes $a[1\pm\frac{1}{2}h_{zz}(\tau)]$.
The reference frame has origin at $\tilde{x}=\tilde{y}=\tilde{z}=0$ and in the coordinates of TT gauge $x=y=0$ and $z=0$ and $t=\tau+O(h)$. 

Recall  that we take real part of $h^{TT}_{zz}$(\ref{eq:PlaneW}), $
h_{zz}(\tau)={\rm Re} \left\{ A_{+}e^{-i\omega\tau}\right\}=({\rm Re}\, A_{+})\cos\omega\tau-({\rm Im} A_{+})\sin\omega\tau$,
where we observe that $\rm{Im}\, A_{+}=0$ then $h_{zz}(\tau)=A_{+}\cos\omega\tau$. 
The laser generated GW resembles a solitary pulse, it can be considered as a superposition of the GW with different frequencies produced in the process. 

The semi-minor axes are $a[1 \pm A\sin\omega\tau]$,
where $A=\frac{1}{2}A_{+}$ and
\begin{align}\label{eq:Ampl}
A\equiv A|_{\tilde{x}_{A}^{j}=0}&=\frac{3}{r}\frac{G}{c^4}S\rho_{0}v_{s}^2(-v_{s}\tau),\nonumber\\
\frac{1}{2}h_{zz}^{TT}(\tau)|_{\tilde{x}_{A}^{j}=0}&=\frac{3}{r}\frac{G}{c^4\omega}S\rho_{0}v_{s}^2(-v_{s}\omega\tau)\sin{\omega\tau}.
\end{align}

\begin{figure}
\centering
\subfigure[The test particles at $\tau=0$, $\tau=2 \pi/\omega$ and more.]{%
\label{fig:Tau0}
\includegraphics[width=0.21\textwidth]{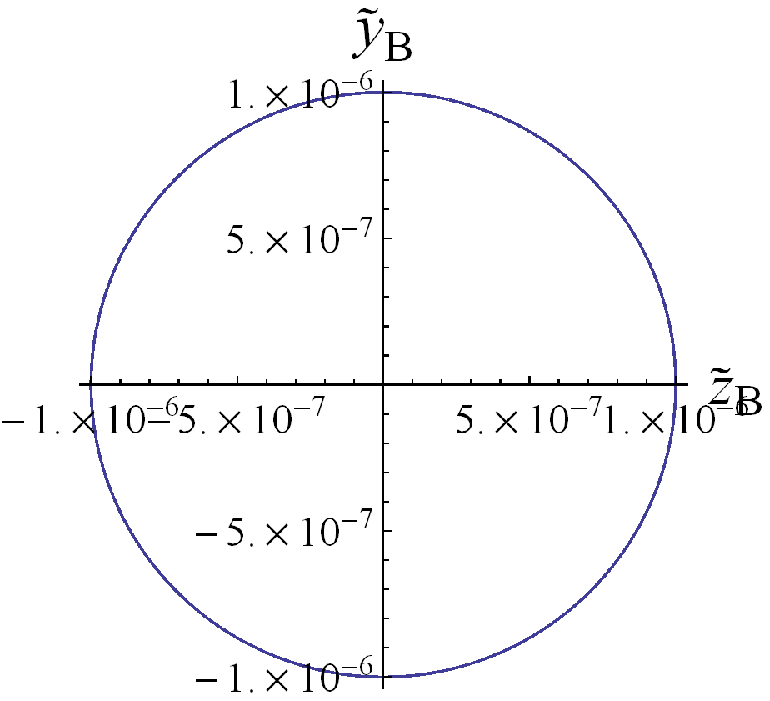}
}
\subfigure[The test particles at $\tau=\pi/2\omega$, $5\pi/2\omega$, $9\pi/2\omega$, $13\pi/2\omega$ and more.]{%
\label{fig:Tau1}
\includegraphics[width=0.19\textwidth]{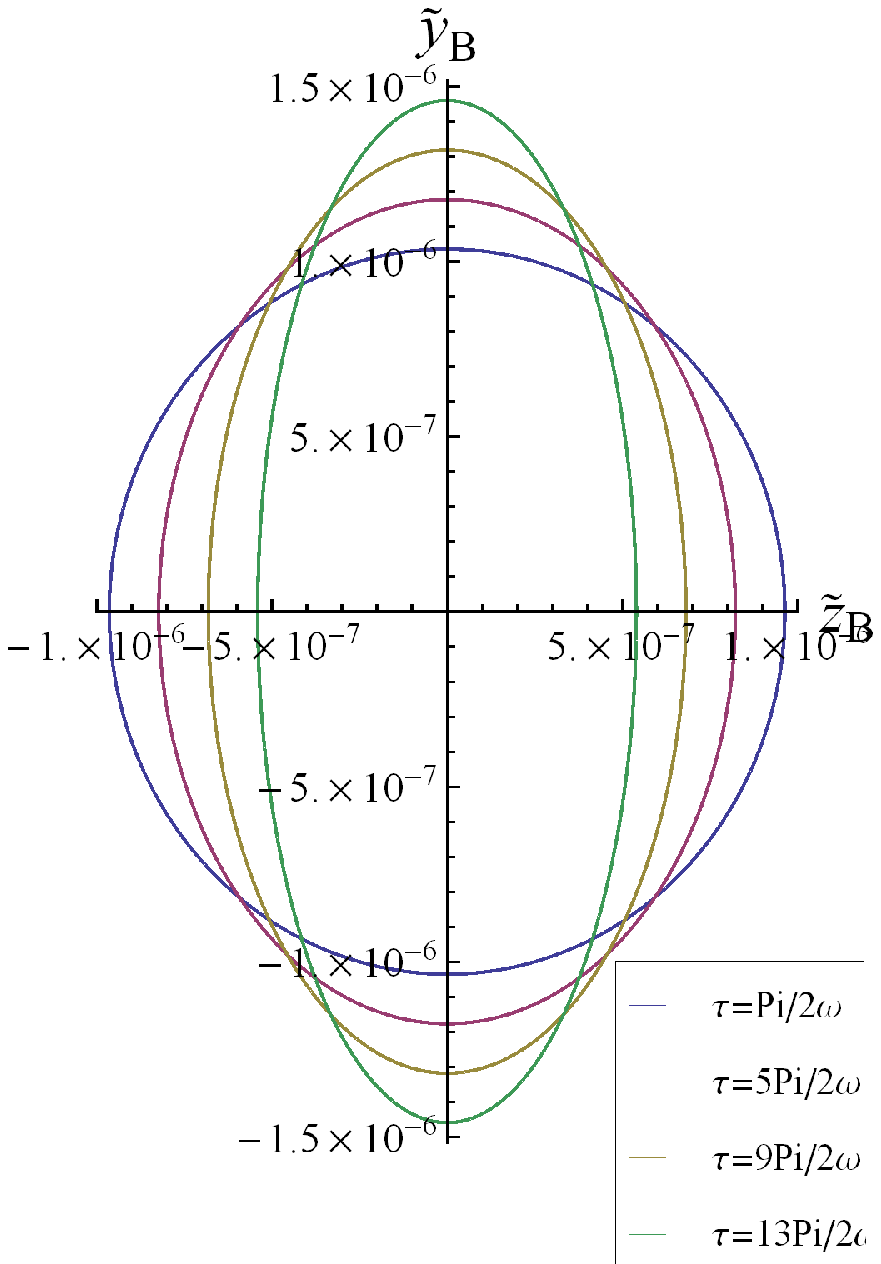}
}\\%
\subfigure[The test particles at $\tau=\pi/\omega$, $3\pi/\omega$, $5\pi/\omega$ and more.]{%
\label{fig:Tau2}
\includegraphics[width=0.21\textwidth]{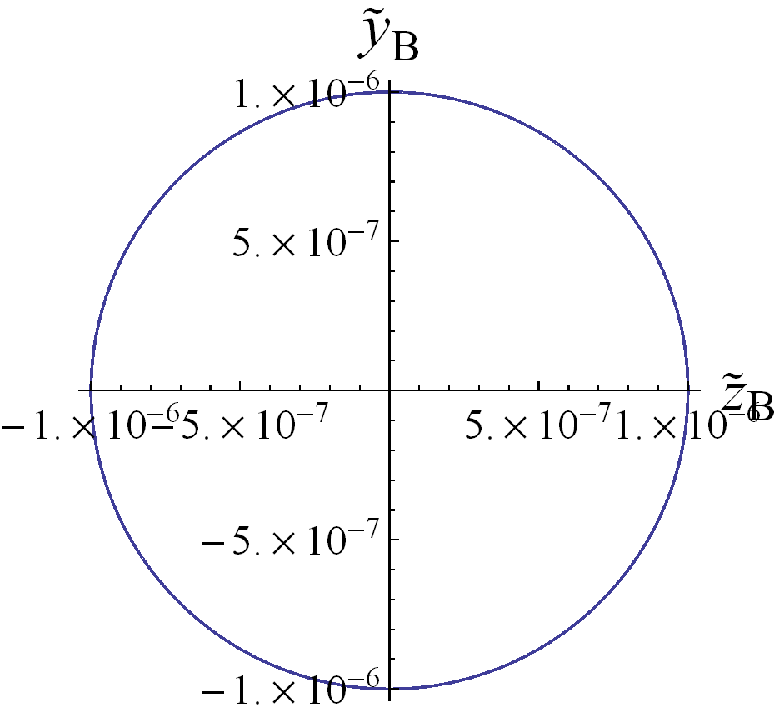}
}
\subfigure[The test particles at $\tau=3\pi/2\omega$, $7\pi/2\omega$, $11\pi/2\omega$, $15\pi/2\omega$ and more.]{%
\label{fig:Tau3}
\includegraphics[width=0.25\textwidth]{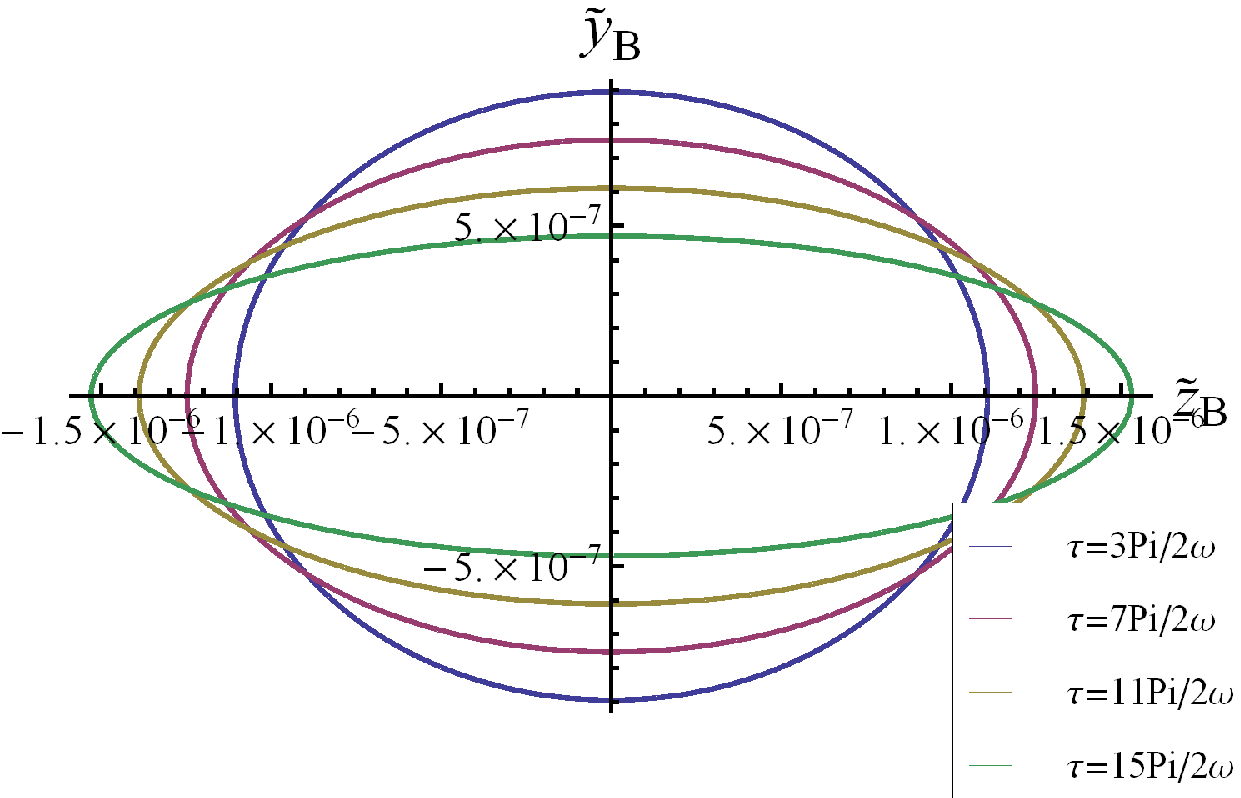}
}%
\caption{\label{fig:TestParticles} (Color online) The diagrams depict the position of test particles as function of time under influence of GW wave with $+$ polarization.}
\end{figure}

The initial position of test particles on circle is pictured in Fig.~\ref{fig:Tau0}  at $\tau=0$, then the circle is gradually changing into ellipse Fig.~\ref{fig:Tau1} due to the influence of the wave, into the direction of $\tilde{y}_{B}$, then it changes to back to circle Fig.~\ref{fig:Tau2}, then again into prolonged ellipse $\tilde{z}_{B}$ Fig.~\ref{fig:Tau3} and then it gets back to circle (Fig.~\ref{fig:Tau0}). 
In our specific case, the ellipse is changing its shape due to the time dependency of the amplitude $A_{+}$.

In Figs.~\ref{fig:Tau1} and \ref{fig:Tau3}, the time dependence of the amplitude is demonstrated explicitly. In Fig.~\ref{fig:Tau1}, we observe the changing of the ellipse to a sharper profile as the time grows, 
in Fig.~\ref{fig:Tau3} the ellipse gets sharper profile in the transversal direction.

For demonstrational purposes, we have used $\frac{1}{2}h_{zz}^{TT}(\tau)|_{\tilde{x}_{A}^{j}=0}=-0.45\,\omega\tau\sin{\omega\tau}$ to make the effect of time dependent amplitude visible. For specific values presented in Section~\ref{subsec:estimate} the expected deformation is of the order of $10^{-39}$ according to Eq.~(\ref{eq:FinalResults}).

\begin{figure}
\centering
\subfigure[The test particles at $\tau=0$, $\tau=2 \pi/\omega$ and more.]{%
\label{Tau0+}
\includegraphics[width=0.23\textwidth]{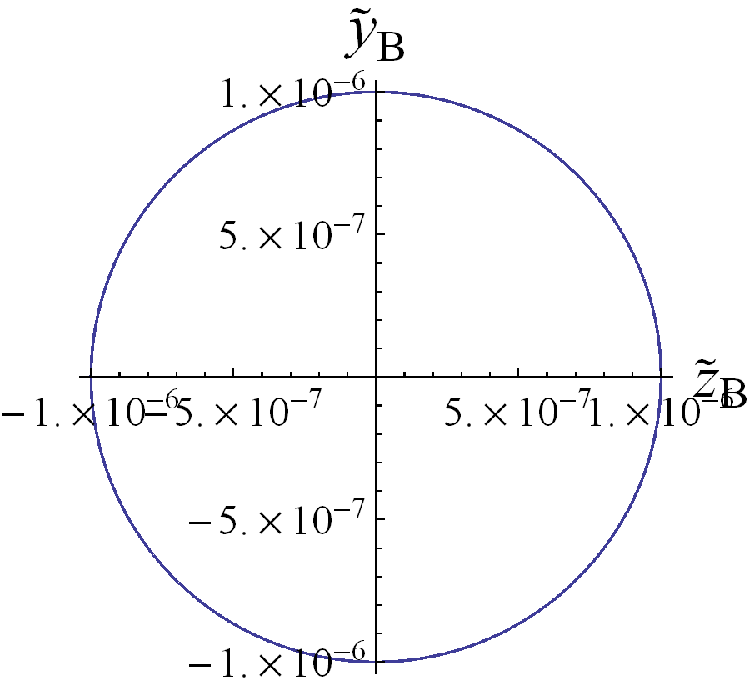}
}
\subfigure[The test particles at $\tau=\pi/2\omega; 5\pi/2\omega; 9\pi/2\omega; 13\pi/2\omega$ and more.]{%
\label{Tau1+}
\includegraphics[width=0.22\textwidth]{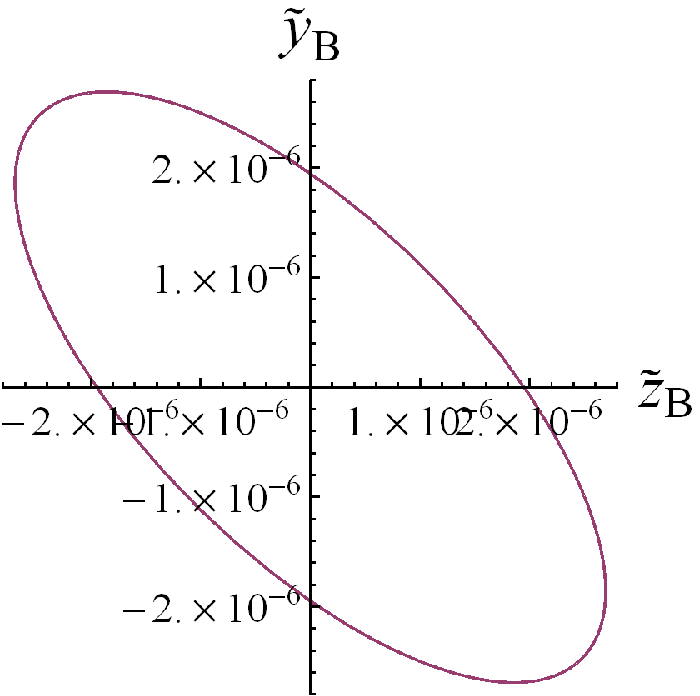}
}\\%
\subfigure[The test particles at $\tau=\pi/\omega;3\pi/\omega; 5\pi/\omega$ and more.]{%
\label{Tau2+}
\includegraphics[width=0.23\textwidth]{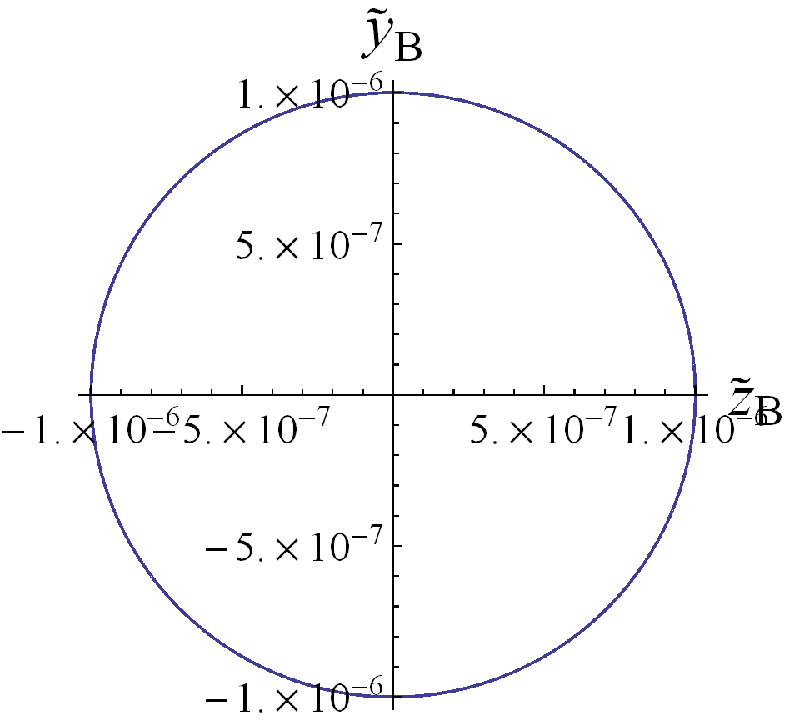}
}
\subfigure[The test particles at $\tau=3\pi/2\omega; 7\pi/2/\omega; 11\pi/2/\omega$,$15\pi/2\omega$ and more.]{%
\label{Tau3+}
\includegraphics[width=0.22\textwidth]{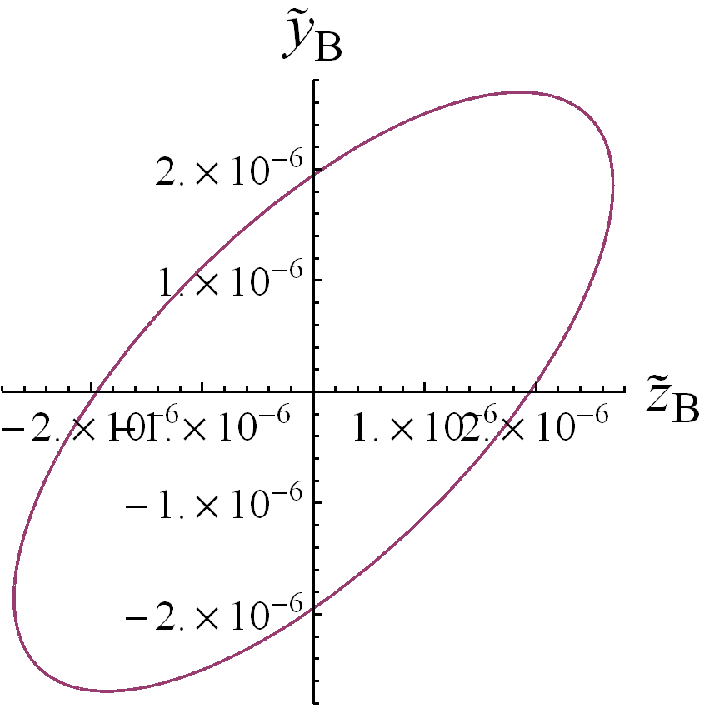}
}\\%
\caption{\label{fig:TestPartSecondPolar} (Color online) The diagrams depict the position of test particles as function of time under influence of GW wave with $\times$ polarization.}
\end{figure}

Similar estimate for the mode $\times$ shows that the deformation of a circle is due to the only non--zero component $h^{TT}_{zy}$. Its amplitude reads $\frac{1}{2}h_{yz}^{TT}(\tau)|_{\tilde{x}_{A}^{j}=0}=-\frac{3}{r}\frac{G}{c^4}S\rho_{0}v_{s}^2\sin{\omega\tau}$,
while we have used function $\frac{1}{2}h_{yz}^{TT}(\tau)|_{\tilde{x}_{A}^{j}=0}=-5\,\sin{\omega\tau}$ for plotting the circle deformation in Fig.~\ref{fig:TestPartSecondPolar}.
The component $h_{yz}^{TT}$ has constant amplitude therefore the ellipses do not change shape when time grows. 
The difference in the ellipses shapes in Figs.~\ref{fig:TestParticles} and \ref{fig:TestPartSecondPolar} shows the possibility to distinguish the wave polarization.

\section{\label{sec:number7} Conclusion}
We have investigated models for the generation of gravitational waves with high energy lasers in laboratory conditions due to excitation of a strong shock.

We demonstrated that a linear gravity can be used for the shock wave model,  calculated and analyzed the perturbation tensor $h^{TT}_{ij}$ and the luminosity of gravitational radiation ${\mathcal{L}}_{GW}$ in low velocity approximation far away from the source. The estimates (\ref{eq:FinalResults}) presented for megajoule class lasers are in agreement with previous publications. 
The results presented in \cite{grossmannMeet2009l,izestELINP2014} are generalized to include the dependence on the laser wavelength and material of the foil.

GW with two independent polarizations + and x modes can be generated. 
The mode $+$ is dipole-like with the maximum emission in the transverse plane. The mode x shows an azimuthal dependence with the maxima emission in the direction is the diagonals of the target. 

The intensity of emitted wave decreases with the distance from the source of radiation as $1/r^2$. These observations could help with the positioning of the detectors in the experiment.

Then the influence of gravitational waves  was analyzed on test particles using the geodesic equation. The difference in amplitudes of the two modes of polarization was demonstrated and visualized in Figs.~\ref{fig:TestParticles} and \ref{fig:TestPartSecondPolar}. The time dependent amplitude of polarization $+$ influences the circle of particles to change the shape to ellipse and as time is passing the ellipse shape becomes sharper. The $\times$ polarization of the GW influences the circle of test particles to change the shape to ellipse shifted by $45^{\circ}$ and the shape stays constant in time. The changes in the shape of the test particles could be in principle measurable in an experiment.


The major problem in the laboratory experiment is the detection of the gravitational waves with the amplitude of the metric perturbation around $10^{-40}$ which is almost $20$ orders lower than the detected radiation from space \cite{AbbottGrav}. The frequencies are in the GHz range, they cannot be detected by any of the known detectors like interferometers or the Weber resonators. The suggested Li--Baker detector is one candidate for the detection of high frequency waves $(10\; {\rm GHz})$ because it uses a different technology. 
Further improvement of detectors in higher frequency range might enable the measurement of the gravitational waves produced in laboratory experiments.

\begin{acknowledgement}
H. Kadlecov\'{a} wishes to thank Tom\'{a}\v{s} Pech\'{a}\v{c}ek for many valuable discussions and time, and Otakar Sv\'{i}tek for his valuable comments. 
The work was supported by the ELI project No. CZ.02.1.01./0.0/0.0/15-008/0000162.
\end{acknowledgement}

%
%






\end{document}